\documentclass[prl,twocolumn,showpacs,amsmath,amssymb]{revtex4}
\usepackage{mathptmx,times,graphicx}

\catcode`\@ 11

\newcommand\partialderiv[3][]{\frac{\partial^{#1}#2}{\partial {#3}^{#1}}}
\def\[{\begin{equation}}
\def\]{\end{equation}}
\def\bse{\begin{subequations}}
\def\ese{\end{subequations}}
\let\<=\langle
\let\>=\rangle
\let\^=\hat
\let\~=\mathbf
\let\==\bar
\let\_=\underline

\let\@trueint=\int
\let\@truesum=\sum
\def\int{\mathop{\textstyle\@trueint}}
\def\sum{\mathop{\textstyle\@truesum}}
\def\half{{\textstyle\frac12}}
\def\dbar{{\bar d}}
\def\E{{\mathbb E}}

\def\d{{\mathrm{d}}}
\def\e{{\mathrm{e}}}
\def\circ{\ifmmode\mathchar"220E\else$\mathchar"220E$\fi}

\def\Re{\mathop{\rm Re}\nolimits}

\def\var{\mathop{\rm var}\nolimits}
\let\trueint=\int
\let\trueiint=\iint
\let\trueiiint=\iiint
\def\int{\mathop{\textstyle\trueint}\limits}
\def\iint{\mathop{\textstyle\trueiint}\limits}
\def\iiint{\mathop{\textstyle\trueiiint}\limits}
\let\eref=\eqref
\let\eqref=\undefined
\advance\parskip0.2pt plus 0.2pt minus 0.2pt
\advance\abovedisplayskip-2pt
\advance\belowdisplayskip-2pt

\begin{document}
\title{Phase noise of dispersion-managed solitons}
\author{Elaine T. Spiller$^{1,2}$ and Gino Biondini$^1$}
\email{biondini@buffalo.edu.}
\affiliation{$^1$ State University of New York at Buffalo, Department of Mathematics,
Buffalo, NY 14260 - USA\\
$^2$ Marquette University, Department of Mathematics Statistics and Computer Science, Milwaukee WI, 53201 - USA}
\begin{abstract}
We quantify noise-induced phase deviations of dispersion-managed solitons
(DMS) in optical fiber communications and femtosecond lasers.
We first develop a perturbation theory for the dispersion-managed
nonlinear Schr\"odinger equation (DMNLSE) in order to compute the
noise-induced mean and variance of the soliton parameters.
We then use the analytical results to guide importance-sampled Monte-Carlo 
simulations of the noise-driven DMNLSE.
Comparison of these results with those from the original,
un-averaged, governing equations confirm the validity of the DMNLSE 
as a model for many dispersion-managed systems, 
and quantify the increased robustness of DMS with respect
to noise-induced phase jitter.
\end{abstract}
\date\today
\maketitle


The performance of many lightwave systems is ultimately limited by 
quantum noise.
Scientifically and technologically important examples include 
optical fiber communication systems and femtosecond (fs) lasers:
the former are a key enabling technology for the information age, 
while Ti:sapphire fs lasers have applications to optical atomic clocks.
Estimating the performance of these systems is a timely problem.
Because both kinds of systems are designed to operate 
with very high accuracies, however, failures result from the
occurrence of unusually large deviations,
which makes calculating error rates extremely difficult.
Direct Monte-Carlo (MC) computations of failure rates are impractical 
due to the exceeding number of samples necessary to obtain 
reliable estimates, 
and analytical predictions are impossible due to the 
scale and complexity of these systems.
In particular, errors in both systems 
are often phase-sensitive,
and both systems employ the technique of dispersion management,
whereby pulses propagate through a periodic concatenation of
components with opposite signs of dispersion 
\cite{MollenauerGordon,YeCundiff}.
%
The probability of rare events can often be efficiently calculated
using importance sampling (IS), with which the noise is sampled
from a biased distribution that makes the rare events occur 
more frequently.
For IS to be successful, however, one must bias towards the
most likely noise realizations that lead to the events of interest.  
For systems modeled by the nonlinear Schr\"odinger equation (NLSE),
this is made possible using soliton perturbation theory (SPT)
\cite{OL28p105,SIREV50p523,PTL17p1022,PTL18p886},
but this tool is not available in dispersion-managed (DM) systems.
Recently~\cite{PRA75p53818}
we developed a perturbation theory for the dispersion-managed NLSE (DMNLSE) 
that governs the long-term dynamics of DM
optical systems~\cite{OL23p1668,OL29p1808,OL21p327},
and we performed ISMC simulations of the pulse amplitude and frequency.
Here we employ this perturbation theory in order to compute
noise-induced means and variances, 
and we develop IS for the DMNLSE by explicitly formulating and solving the 
optimal biasing problem.
We then perform ISMC simulations of the pulse phase, 
where the choice of biasing is non-trivial.
Finally, we compare these results to the original, un-averaged 
system as well as to systems with constant dispersion.

\paragraph*{Perturbations of dispersion-managed solitons.}

The propagation of optical pulses in dispersion-managed 
fiber communication systems~\cite{MollenauerGordon} and 
Ti:sapphire lasers~\cite{YeCundiff} 
is described by an equation which we refer to as NLSE+DM:
\[
i\partialderiv qz+\half\,d(z/z_a)\,\partialderiv[2]qt+g(z/z_a)|q|^2q=
  i\nu(t,z)\,.
\label{e:NLS}
\]
Here $z$ is the propagation distance, $t$ is the retarded time,
$q(t,z)$ is the slowly varying electric field envelope (rescaled to
account for loss/amplification), $d(z/z_a)$ is the local dispersion,
and $g(z/z_a)$ describes the periodic power variation due to
loss/amplification. The choice of $d(z/z_a)$ is called a dispersion
map, and $z_a$ is the dispersion map period. 
The forcing is 
$\nu(t,z)=
\sum\nolimits_{n=1}^{N_a} \nu_n(t)\,\delta(z-nz_a)\,$,
where
$\delta(z)$ is the Dirac delta and $\nu_n(t)$ is 
white Gaussian noise, satisfying $\E[\nu_n(t)]=0$ and
$\E[\nu_n(t)\nu^*_{n'}(t')]= \sigma\delta(t-t')\,\delta_{nn'}$, 
where $\E[\,\cdot\,]$ denotes ensemble average,
the asterisk complex conjugation, 
$\delta_{nn'}$ is the Kronecker delta 
and $\sigma^2$ is the noise variance. 

Once the compression/expansion of the pulse in each dispersion map
is properly factored out, the core pulse shape 
obeys the DMNLSE \cite{OL23p1668,OL21p327}.   
Namely, to leading order we can approximate the solution 
of Eq.~\eref{e:NLS} as
$\^q(\omega,z,\zeta)= \e^{-iC(\zeta)\omega^2/2}\^u(\omega,z)$,
where 
$\^f(\omega)= \int \e^{-i\omega t}f(t)\,\d t$ 
is the Fourier transform of $f(t)$
(all integrals are complete unless limits are given),
and $\zeta=z/z_a$. 
Here $C(\zeta)= z_a\int\nolimits_0^\zeta\big(d(\zeta')-\dbar\big)\,\d\zeta'\,$,
where $\dbar$ is the average dispersion.
The exponential factor in front of $\^u(\omega,z)$
accounts for the rapid ``breathing'', 
while the slowly varying envelope $\^u(\omega,z)$ 
satisfies the perturbed DMNLSE:
\begin{multline}
\kern-0.4em
i\partialderiv uz+\half\dbar\partialderiv[2]ut
\\ 
  +\iint u^{}_{(t+t')}u^{}_{(t+t'')}u^*_{(t+t'+t'')}R^{}_{(t',t'')}\,\,\d t'\d t'' 
     = i\nu(t,z)\,,\kern-0.4em
\label{e:DMNLS}
\end{multline}
where the asterisk denotes complex conjugate, 
and for brevity $u_{(t)}=u(t,z)$, 
etc. 
The kernel $R(t',t'')$ quantifies the average nonlinearity over
a dispersion map mitigated by dispersion management:
$R(t't'')= \iint \e^{i\omega't'+i\omega''t''}
  r(\omega'\omega'')\d\omega'\d\omega''$,
where $r(x)= \int\nolimits_0^1 \e^{ix C(\zeta)}g(\zeta)\,\d\zeta$.

The DMNLSE and
its solutions depend on a parameter~$s$, called the 
\textit{reduced map strength}, which quantifies the size of the 
dispersion variations around their mean.
Dispersion-managed solitons (DMS) are traveling-wave solutions of the DMNLSE. 
If $u_o(t,z;s)= \e^{i\lambda^2z/2}f(t;s)\,$, then
$\^f(\omega)$ satisfies a nonlinear integral equation which can be
efficiently solved numerically \cite{PRA75p53818}. 
The invariances of the DMNLSE then yield from $u_o(t,z;s)$ a
four-parameter family of DMS:
\[
u_{\mathrm{dms}}(t,z;s)= \e^{i\Theta(t,z)}A\,f(A(t-T);A^2s)\,,
\label{e:dms}
\] 
where
$A$ and $\Omega$ are the DMS amplitude and frequency, 
$\Theta(t,z)= \Omega\,(t-T)+\Phi$ is the local phase, and 
$T$ and $\Phi$ are respectively the mean time
and mean phase.  
In the unperturbed case, the mean time and mean phase
evolve according to
$\dot T=\Omega\,$ and 
$\dot\Phi=(A^2+\Omega^2)/2\,$.
[Hereafter, the dot denotes differentiation with respect to $z$.]

In the presence of perturbations, the DMS will evolve.
If $u(t,z) =u_{\mathrm{dms}}+ w$ solves the
noise-perturbed DMNLSE,
$w(t,z)$ satisfies the corresponding perturbed linearized DMNLSE. 
But part of the noise goes to change the soliton parameters.
The noise-induced DMS parameter changes at each map period 
are found by demanding that $w(t,z)$ remain small, and 
are written as $Q(nz_a^+)= Q(nz_a^-)+\Delta Q\,$, 
where for $Q=A,\Omega,T$ it is
$\Delta Q= \<e^{i\Theta}\_y_Q,\nu_n(t)\>\big/ \<\_y_Q,y_Q\>\,$,
while \cite{PRA75p53818}
\[
\Delta\Phi= \<e^{i\Theta}\_y_\Phi,\nu_n(t)\>\big/ \<\_y_\Phi,y_\Phi\>
  + \Omega\,\<e^{i\Theta}\_y_T,\nu_n(t)\>\big/ \<\_y_\Omega,y_\Omega\>\,.
\label{e:paramchanges}
\]
Here $\<f,g\>= \Re\int\!f^*(t)g(t)\,\d t$ is the inner product,
$y_A(t),\dots,y_\Phi(t)$ are the neutral eigenmodes and 
generalized eigenmodes of the linearized DMNLSE, 
and $\_y_A(t),\dots,\_y_\Phi(t)$ are the adjoint modes
\cite{PRA75p53818}:
$y_T= -\partial U/\partial\xi\,$,
$y_\Omega= i\xi U$,
\[
y_\Phi= iU\,,\quad
y_A= \frac1A\bigg(U + \xi\partialderiv U\xi + 2s\partialderiv Us\bigg)\,,
\label{e:modes}
\]
while $\_y_\Phi= i y_A$,
$\_y_T= - iy_\Omega/A$,
$\_y_\Omega= - y_T/A$,
and $\_y_A= U$.
Here $\xi= t- T(z)$, and
$U(t,z)= u(t,z)\,\e^{-i\Theta}$
is the DMS envelope.
All of these results reduce to those arising from soliton
perturbation theory for the NLSE \cite{SIREV50p523} when $s=0$.

\paragraph*{Noise-induced parameter variances.}

When the perturbation in~\eref{e:DMNLS} represents noise,
the above results yield a system of nonlinear 
stochastic differential equations (SDEs) for the evolution 
of the DMS parameters under the effect of noise:
\begin{gather}
\label{e:SDEs}
\dot Q= \nu_Q(z)\,,\qquad
\dot\Phi = \half(A^2+\Omega^2) + \Omega\,\nu_T(z)+ \nu_\Phi(z)\,,
\end{gather}
for $Q=A,\Omega,T$,
where the source terms are
$\nu_Q(z)= \<\e^{i\Theta}\_y_Q,S\>\big/\<\_y_Q,y_Q\>$ for all $Q$.
We employ a continuum approximation of Eqs.~\eref{e:SDEs}, 
considering
$\nu(t,z)$ to be a zero-mean Gaussian white-noise process 
with autocorrelation
$\E[S(t,z)S^*(t,z)]= \sigma^2\,\delta(t-t')\delta(z-z')\,$.
The sources $\nu_A(z),\dots,\nu_\Phi(z)$ 
are then independent zero-mean white-noise processes, with 
autocorrelation 
$\E[S_Q(z)S_{Q'}(z')]= \sigma_Q^2\delta(z-z')\,$,
where
$\sigma_Q^2= \sigma^2\|\_y_Q\|^2\big/\<\_y_Q,y_Q\>^2\,$.
All of these variances depend on the soliton amplitude $A$
as well as on the map strength~$s$, 
and therefore on the propagation distance~$z$.
As a result, it is not possible to integrate Eqs.~\eref{e:SDEs} 
in closed form, even in the case of constant dispersion.
If the amplitude deviations are not large, one can 
approximate $\sigma_A^2,\dots,\sigma_\Phi^2$ as constant.
In this limit, Eqs.~\eref{e:SDEs} can be integrated exactly,
to give
$Q(z)= Q_o+ W_Q(z)$ for $Q=A,\Omega$, while
\bse
\begin{gather}
T(z)= T_o+ \int_0^z\Omega(z')\d z' + W_T(z)\,,
\\[-1ex]
\Phi(z)= \half\int_0^z\big(A^2(z')+\Omega^2(z')\big)\,\d z'
  + \int_0^z\Omega(z')S_T(z')\,\d z' + W_\Phi(z)\,,
\nonumber\\[-2ex]
\end{gather}
\ese
where
$W_Q(z)= \int\nolimits_0^zS_Q(z')\,\d z$
is a zero-mean Wiener process with autocorrelation 
$\E[W_Q(z)W_{Q'}(z')]= \sigma_Q^2\delta_{QQ'}\min(z,z')\,$.
Unlike other soliton parameters, the mean value of the 
soliton phase is affected by the noise:
\[
\E[\Phi(L)]= \half(A_o^2+\Omega_o^2)\,L+ \frac14(\sigma_A^2+\sigma_\Omega^2)\,L^2\,.
\]
Stochastic calculus also yields the variances of the noise-perturbed 
output soliton parameters as
$\var[A(L)]= \sigma_A^2L$, $\var[\Omega(L)]= \sigma_\Omega^2L$,
$\var[T(L)]= \sigma_T^2L + \frac13\sigma_\Omega^2L^3$ and
\begin{multline}
\var[\Phi(L)]=  (\sigma_\Phi^2+\Omega_o^2\sigma_\Omega^2)\,L
   +\Omega_o\sigma_T^2\sigma_\Omega^2\,L^2
\\
   +\frac13 (A_o^2\sigma_A^2+\Omega_o^2\sigma_\Omega^2)\,L^3
   + \frac1{12} (\sigma_A^4+\sigma_\Omega^4)L^4\,.
\label{e:phivar}
\end{multline}
The cubic dependence on distance of the phase jitter 
due to the Kerr effect 
is the Gordon-Mollenauer jitter \cite{OL15p1351}, 
but note that additional contributions are present.
Remarkably, these results are formally identical to those for 
the NLSE~\cite{JOSAB21p266}.  
The dependence of the variance on the soliton amplitude, however, 
is dramatically different, due to the different dependence on~$A$ 
of the norms and inner products~\cite{PRA75p53818}.
More importantly, these results are not enough to accurately estimate 
the occurrence of those rare events in which the noise produces large 
phase deviations, because:
(i)~the prediction for the mean phase is inaccurate, 
as we show below;
(ii)~the knowledge of noise-induced means and variances is not enough 
to estimate behavior in the tails, because not all soliton parameters 
are Gaussian-distributed; 
(iii) even if the output probability density functions (PDFs) were Gaussian, 
extrapolating the results to reach the distribution tails
would magnify all uncertainties exponentially, 
thereby making any prediction meaningless.

\paragraph*{Most-likely noice-induced phase deviations.}

Even though perturbation theory is not enough by itself
to predict failure rates,
it provides a key tool to implement IS.
To successfully apply IS, one must first find the 
most likely noise realization subject to the constraint of 
achieving a given parameter change. 
For additive white Gaussian noise, this problem is solved by minimizing 
the negative of the argument of the exponential in the noise PDF,
namely the integral $\int |\nu_n(x)|^2dx\,$, subject to the constraint 
$\Delta Q_n={}\Delta Q_{\mathrm{target}}$. 
The solution is~\cite{PRA75p53818}:
\[
\nu_{n,\mathrm{opt}}(t)=
  \Delta Q_{\mathrm{target}}\,\e^{i\Theta(z)} \_y_Q(t)\,
      \<\_y_Q,y_Q\>/\|\_y_Q\|^2\,.
\]
To induce a larger than normal parameter change, one can then bias 
the noise
by concentrating the MC samples around $\nu_{n,\mathrm{opt}}(x)$.
That is, 
$\nu_{n,\mathrm{biased}}(t)=\nu_{n,\mathrm{opt}}(t)+\nu_n(t)$,
where $\nu_{n,\mathrm{opt}}(t)$ is given above
and $\nu_n(t)$ is unbiased.

Once the most likely noise realization that produces a
given parameter change $\Delta Q_n$ at each map period is known,
one must also find the most likely way to distribute
a total parameter change $\Delta Q_{\mathrm{tot}}$ at the output
among all map periods.
In principle, when seeking large phase changes,
one must bias an appropriate combination of all linear modes.
Among the terms in the right-hand-side of Eq.~\eref{e:SDEs}, however,
changes in $\Omega^2$ and $\<\e^{-i\Theta}\_y_T,S\>\Omega$ 
are second-order in the noise, while changes in $A^2$ are
first-order in the noise, because $\Omega_o=0$ while $A_o\ne0$.
We thus introduce the auxiliary quantity $\phi(z)$ such that
$d\phi/dz= A^2/2 + \nu_\Phi(z)$ and $\phi(0)=\Phi_o$, 
and consider the optimal biasing problem for $\phi(z)$.
In the continuum approximation, the biasing function is then
\[
\label{e:biasingcont}
b(t,z)= \dot A\,\_y_A\,{\<y_A,\_y_A\>}/{\|\_y_A\|^2}
  + (\dot\phi-A^2/2)\,\_y_\Phi\,{\<y_\Phi,\_y_\Phi\>}/{\|\_y_\Phi\|^2}\,.
\]
[The direct phase biasing is not given by $\dot\phi\,z_a$,
but rather by $(\dot\phi-A^2/2)\,z_a$.]
Minimizing the sum of the $L_2$ norm of this biasing function 
over all amplifiers is equivalent 
to finding functions $A(z)$ and $\phi(z)$ that minimize 
the functional
\[
J[A,\phi]=
\int_0^L\bigg[\frac1{\sigma_A^2}\dot A^2
  + \frac1{\sigma_\Phi^2}\big(\dot\phi- A^2/2\big)^2\bigg]\,\d z\,.
\label{e:phifunctional}
\]
The Euler-Lagrange equations associated with $J[A,\phi]$ yield
\bse
\label{e:phasebiasing}
\begin{gather}
\dot\phi-A^2/2 = c\,\sigma_\Phi^2\,,
\label{e:dphidzbias}
\\
2\ddot A\frac1{\sigma_A^2}
  +\dot A^2\partialderiv{ }A\bigg[\frac1{\sigma_A^2}\bigg] 
  + c^2 \partialderiv{ }A\big[\sigma_\Phi^2\big] 
  + 2c\,A = 0\,,
\label{e:lambdaODE}
\end{gather}
\ese
where $c$ is a Lagrange multiplier. 
The solution of the system composed of Eqs.~\eref{e:phasebiasing}, together 
with the boundary conditions
$A(0)=A_o$,\, $\phi(0)=\dot A(L)=0$\, and\, 
$\phi(L)= \phi_\mathrm{target}$, 
determines the optimal amplitude and phase paths 
around which one must bias the ISMC simulations.
(The condition $\dot A(L)=0$ 
applies because amplitude changes at $z=L$ 
do not produce phase changes.)
This system can be integrated numerically using relaxation methods
or numerical continuation software.
Different output phases can be targeted by 
solving the system for different values of~$c$,
which determines the amount of biasing being applied
($c=0$ yields no bias).
Equations~(\ref{e:biasingcont}--\ref{e:phasebiasing}),
which are new, 
reduce to known results
in the case of constant dispersion \cite{SIREV50p523,Spiller}.
But unlike the constant-dispersion case (and unlike the case of time biasing),
here the direct phase biasing is not constant in~$z$.
Physically, this is a consequence of the different way in which 
noise is translated into phase jitter in the DMNLSE by way of
the linear modes.

\paragraph*{ISMC simulations.}

We now discuss importance-sampled MC (ISMC) simulations 
aimed at computing the PDF 
of the soliton phase at the output. 
To quantify larger-than-normal phase deviations,
we perform the following steps at each map period:\,\  
(i)~recover the underlying DMS from the noisy signal;\,\  
(ii)~obtain the linear modes and adjoint modes of the linearized DMNLSE
    around the given DMS;\,\ 
(iii)~generate an unbiased noise realization, shift its mean with
the appropriately scaled adjoint modes and update the 
likelihood ratios~\cite{PRA75p53818}. 
We then add the noise to the pulse,
propagate the noisy signal to the next map period, 
and repeat this process until the signal reaches the output. 
For each noise realization,
the full DMNLSE is used to propagate the signal.
(The linearized DMNLSE is only used to guide IS via its modes.)
Even though the noise-induced DMS parameter changes at each map period 
are small, the accumulation of these changes often results in 
a significantly distorted output signal.

We choose system parameters based on realistic values for optical
fiber communications. 
Typical values of system parameters for fs lasers 
can be obtained from Ref.~\cite{PRL94p243904}.\,\  
We consider a piecewise constant dispersion map, with equal-length
normal and anomalous dispersion sections and local dispersion 
coefficients of 23.27 and $-$22.97\,ps$^2$/km,
resulting in an average dispersion of 0.15\,ps$^2$/km.
We set the unit time to 17\,ps, corresponding to $\=d=1$
and $s=4$,
and we use the resulting dispersion length of 1,923\,km 
to normalize distances.
We consider a transmission distance of 6000\,km (or $L=3.1201$) 
and amplifiers spaced 100\,km apart,
for a total of $N_a=60$ 
with dimensionless spacing~$z_a=0.052$,
and we set the map period to be aligned with them.
We take a nonlinear coefficient of 1.7\,(W$\cdot$km)$^{-1}$,
a peak power of 3.51\,mW,
a loss coefficient of 0.25\,dB/km,
a spontaneous emission factor of 1.65,
resulting in a dimensionless noise variance $\sigma^2=1.873\cdot10^{-3}$
and an OSNR of 9.3\,dB.
We normalize pulse powers with the power needed to have $\=g=1$, 
namely 3.51\,mW.
Finally, we take input pulses to have unit peak amplitude, 
resulting in $A_o=1$.

\begin{figure}[t!]
\centerline{\includegraphics[width=0.405\textwidth]{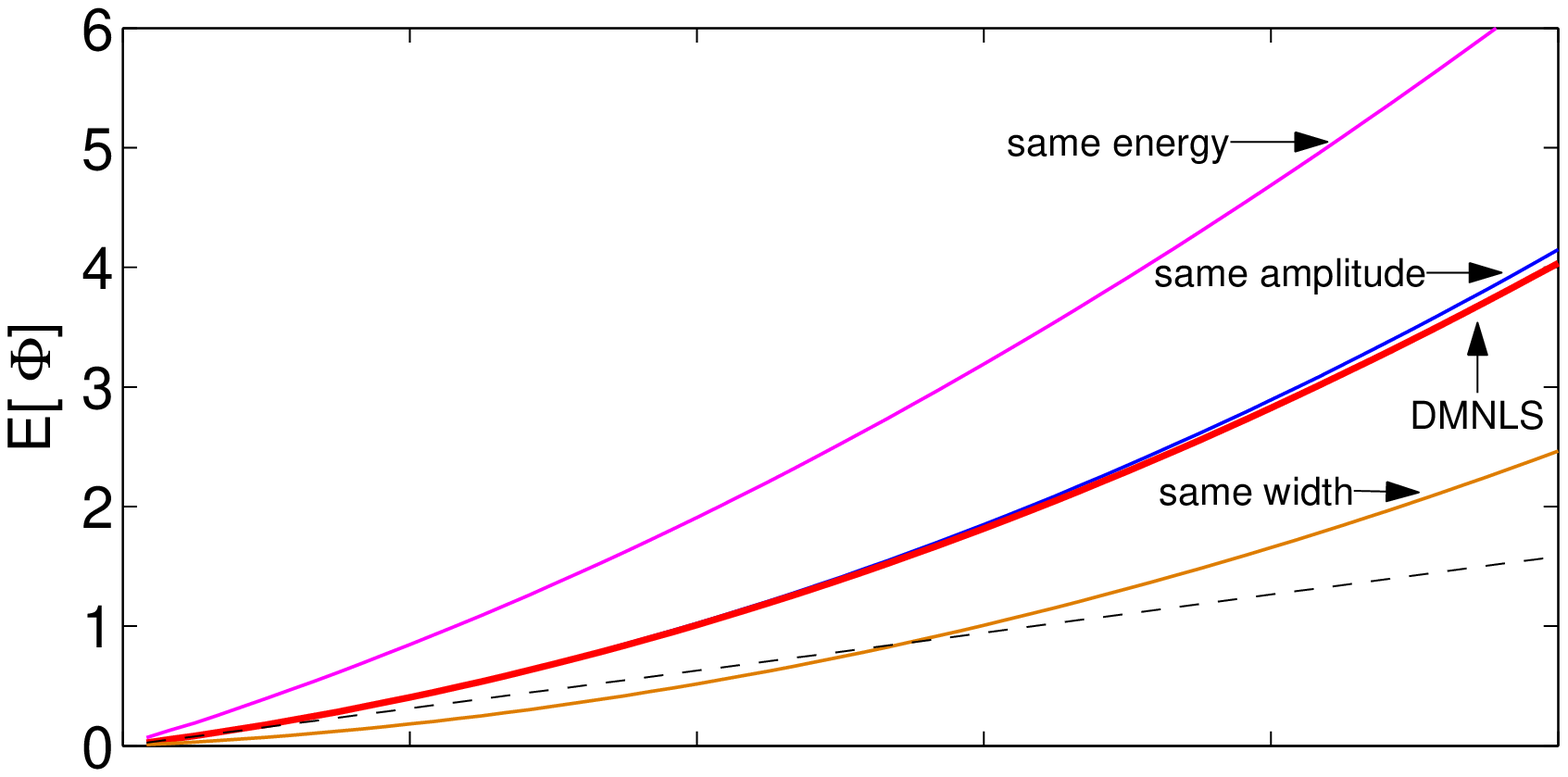}\quad}
\kern-\smallskipamount
\centerline{\includegraphics[width=0.405\textwidth]{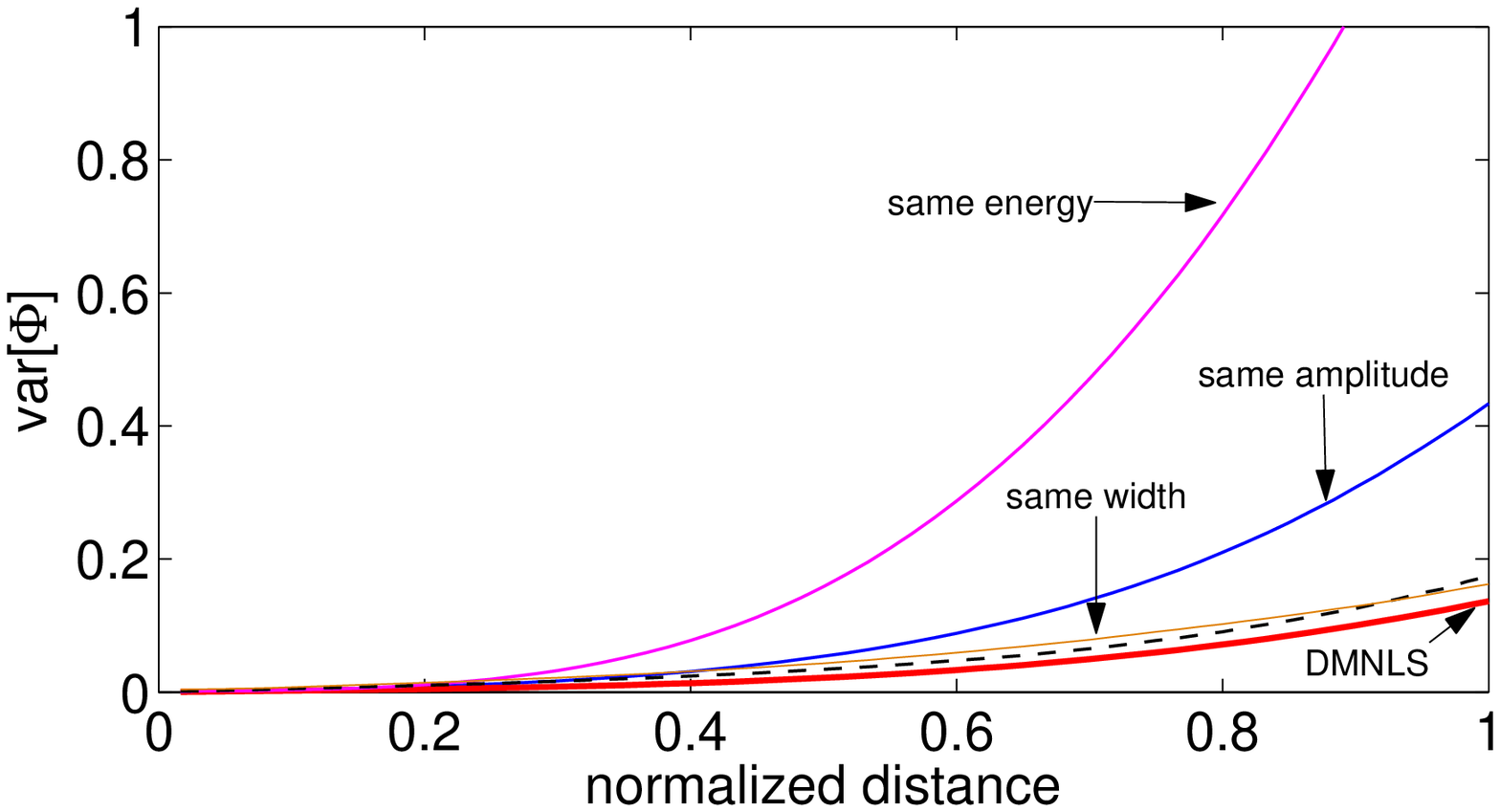}\quad}
\kern-\smallskipamount
\caption{(Color online) Mean (top) and variance (bottom) of the soliton phase as reconstructed
with standard MC simulations.  
Thick red lines: DMNLSE;
blue lines: constant-dispersion NLSE soliton with same amplitude as the DMS;
magenta lines: NLSE soliton with same energy;
orange lines: NLSE soliton with same width.
The dashed black lines show the predictions from perturbation theory.}
\label{f:phasemeanvar}
\medskip
\leftline{\kern1em\includegraphics[width=0.415\textwidth]{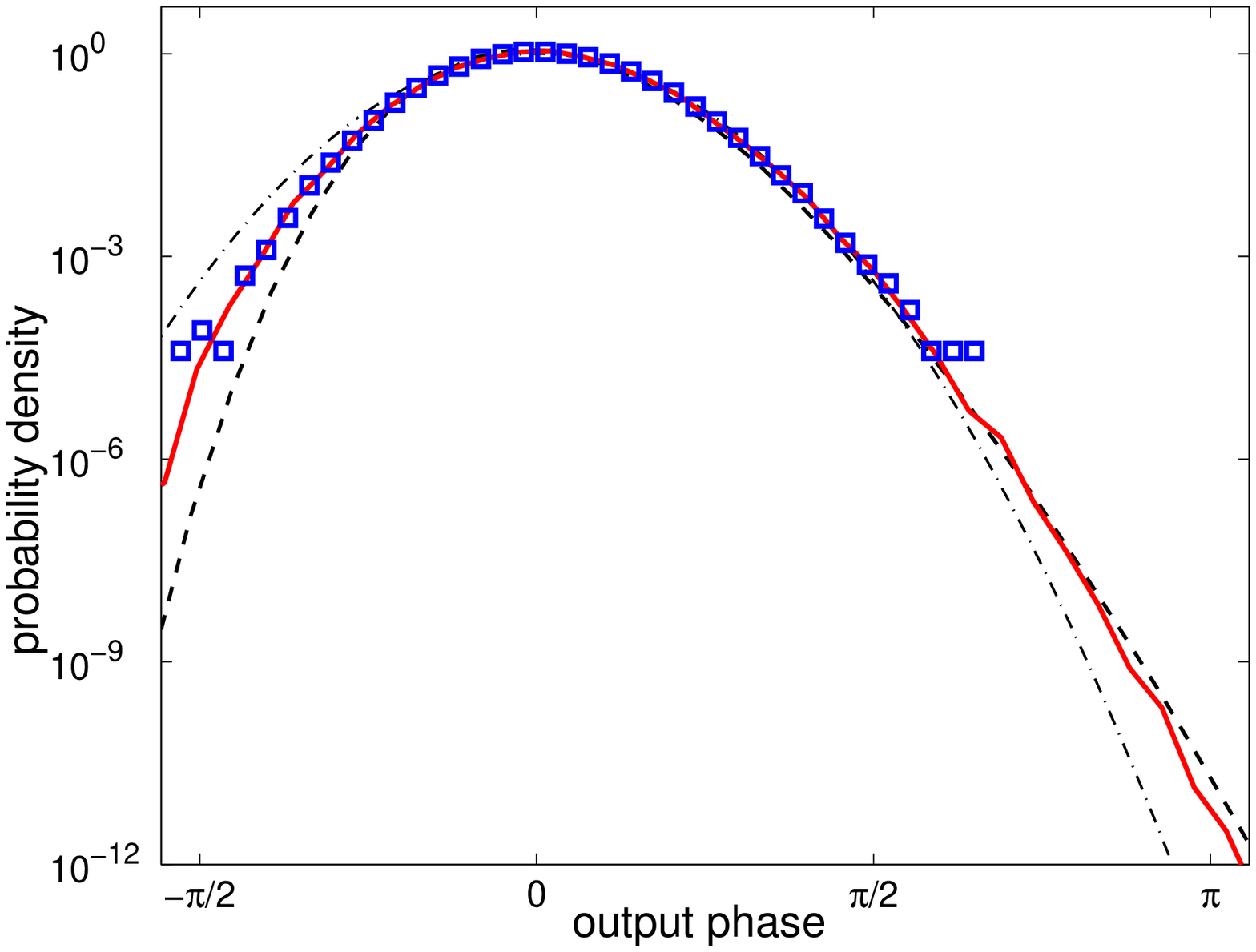}}
\kern-\smallskipamount
\caption{(Color online) PDF of output phase. Thick red curve: 
ISMC simulation of DMNLSE with 50,000 samples. 
Blue squares: standard MC simulation of the NLSE+DM with 250,000 samples.
Dashed curve: ISMC simulation of the noise-driven SDEs \eref{e:SDEs}.
Dot-dashed curve: a Gaussian PDF with variance given by~\eref{e:phivar}.}
\label{f:phasepdf}
\kern-\medskipamount
\end{figure}

\paragraph{Numerical results and discussion.}
 
Figure~\ref{f:phasemeanvar} shows the numerically reconstructed phase mean 
and variance as a function of distance for the DMNLSE, as well as 
the corresponding values for constant-dispersion NLSE solitons 
with same mean, amplitude or energy as the DMS, plus the 
predictions of perturbation theory.
Note that the DMS has the lowest variance of all.
(The constant-dispersion NLSE soliton with same width as the DMS 
has a much lower energy, which makes it much more susceptible to
Gordon-Haus jitter.)
Note that the means and variances of the 
numerically reconstructed phase 
depend dramatically on the particular definition of phase used
in the simulations.
Hence, consistency is crucial to ensure agreement between theory
and simulations.
Here, the phase of a noisy pulse is defined (both in the theory and 
in the numerics) as that of the underlying DMS (obtained as in 
\cite{PRA75p53818}).

A significant discrepancy is evident between analytical and 
numerical results for the mean phase.
No satisfactory explanation currently exists for this effect, 
which also occurs for the NLSE \cite{Spiller}.
It is likely to depend on a failure of SPT and/or from 
second-order effects.
(Numerical results show that the discrepancy also depends on the
computational noise bandwidth.)
On the other hand, the analytical prediction for the variance agrees 
very well with the numerical results, both for NLSE and DMLNSE.

Figure~\ref{f:phasepdf} shows the PDF of the DMS output phase
as computed from ISMC simulations of the DMNLSE~\eref{e:DMNLS},
standard MC simulations of the NLSE+DM~\eref{e:NLS}, 
plus a Gaussian distribution with variance given by perturbation theory
and a PDF obtained from direct ISMC simulations of the SDEs~\eref{e:SDEs}. 
The ISMC results collect samples generated with a few biasing targets, 
using multiple IS \cite{JLT22p1201} to properly combine the data.
The PDFs from both the DMNLSE and the NLSE clearly deviate from Gaussian, 
but they agree very well with each other as far down in probability 
as the unbiased MC simulations can reach. 
Conversely, while the Gaussian approximation agrees well near the peak of 
the PDF, for deviations from the mean phase of $\pi$ or more
(a value that is relevant for fs lasers)
it is off by several orders of magnitude.
Similarly, the SDEs obtained from perturbation theory fail to
accurately reproduce the full dynamics of the soliton phase
at lower-than-average values of phase.
Remarkably, however, ISMC simulations guided by perturbation theory 
yield the correct phase behavior.


Importantly, results from the noise-perturbed DMNLSE and NLSE+DM 
agree \textit{pathwise}, not just in the overall PDFs at the output.
That is, they agree for each noise realization as a function of distance.
These results, which are surprising given the ``softness'' of the phase
and the complexity of the system 
(nonlinearity, dispersion, noise, large deviations etc.),
provide further confirmation of the validity and robustness of the DMNLSE 
in capturing the essential dynamics of DM systems.  
Its usefulness is also increased by the availability of tools such as 
the perturbation theory 
presented here, an analogue of which is lacking for the NLSE+DM.

To our knowledge, 
this is the first time that the probability of large phase deviations 
in dispersion-managed systems has been quantified.
Similar dynamics should also arise for non-solitonic pulses,
but the analysis for that case will be more complicated because 
generic pulses do not preserve a flat phase across their 
temporal profile upon propagation.

An important question is also whether these results can be used in fs lasers
in order to quantify the probability of the occurrence of phase slips 
in optical atomic clocks.
Since gain and loss play an obvious role in lasers,
one could expect that it will be necessary to derive a perturbation theory 
for the non-conservative version of the DMNLSE that was derived 
as a model for fs lasers \cite{NLTY21p2849}.
Since the DMNLSE itself provides a surprisingly good quantitative description 
of these lasers \cite{PRL94p243904}, however, 
whether or not such an extension will indeed be necessary 
remains at present an open question.

\begin{acknowledgments}
We thank S.~T.\ Cundiff and W.~L.\ Kath for many insightful discussions.
This work was partially supported by NSF under grants DMS-0112069,
DMS-0506101 and DMS-0757527.
\end{acknowledgments}
\par\kern-2\medskipamount

\catcode`\@ 11
\def\journal#1&#2,#3(#4){\begingroup{\sl #1\unskip}~{\bf\ignorespaces #2}\rm, #3 (#4)\endgroup}
\def\title#1{\textit{#1}}

\vglue-0.4\smallskipamount
\end{document}